\documentclass[aps,prl,twocolumn,showpacs,superscriptaddress,showkeys]{revtex4-1}
\usepackage{graphicx,color}
\usepackage{hyperref}

\begin{document}
\title{Experimental Observation of a Fundamental Length Scale of Waves in Random Media}

\author{S.~Barkhofen}
\affiliation{Fachbereich Physik, Philipps-Universit\"{a}t Marburg, Renthof 5, 35032 Marburg, Germany}
\author{J. Metzger}
\affiliation{Max-Planck-Institute for Dynamics and Self-Organization (MPIDS), 37077 Goettingen, Germany}
\affiliation{Institute for Nonlinear Dynamics, Department of Physics, University of G\"{o}ttingen, 37077 G\"{o}ttingen, Germany}
\author{R. Fleischmann}
\affiliation{Max-Planck-Institute for Dynamics and Self-Organization (MPIDS), 37077 Goettingen, Germany}
\author{U.~Kuhl}
\affiliation{LPMC, CNRS UMR 7336, Universit\'{e} de Nice Sophia-Antipolis, F-06108 Nice, France}
\email{ulrich.kuhl@unice.fr}
\affiliation{Fachbereich Physik, Philipps-Universit\"{a}t Marburg, Renthof 5, 35032 Marburg, Germany}
\author{H.-J.~St\"{o}ckmann}
\affiliation{Fachbereich Physik, Philipps-Universit\"{a}t Marburg, Renthof 5, 35032 Marburg, Germany}
%PACS: 42.25.Dd Wave propagation in random media
%PACS: 03.65.Sq Semiclassical theories and applications
%PACS: 05.45.Mt Quantum chaos; semiclassical methods

\begin{abstract}
Waves propagating through a weakly scattering random medium show a pronounced branching of the flow accompanied by the formation of freak waves, i.e., extremely intense waves. Theory predicts that this strong fluctuation regime is accompanied by its own fundamental length scale of transport in random media, parametrically different from the mean free path or the localization length. We show numerically how the scintillation index can be used to assess the scaling behavior of the branching length. We report the experimental observation of this scaling using microwave transport experiments in quasi-two-dimensional resonators with randomly distributed weak scatterers. Remarkably, the scaling range extends much further than expected from random caustics statistics.
\end{abstract}

\maketitle

\begin{figure}[ht!]
\includegraphics[width=0.90\columnwidth]{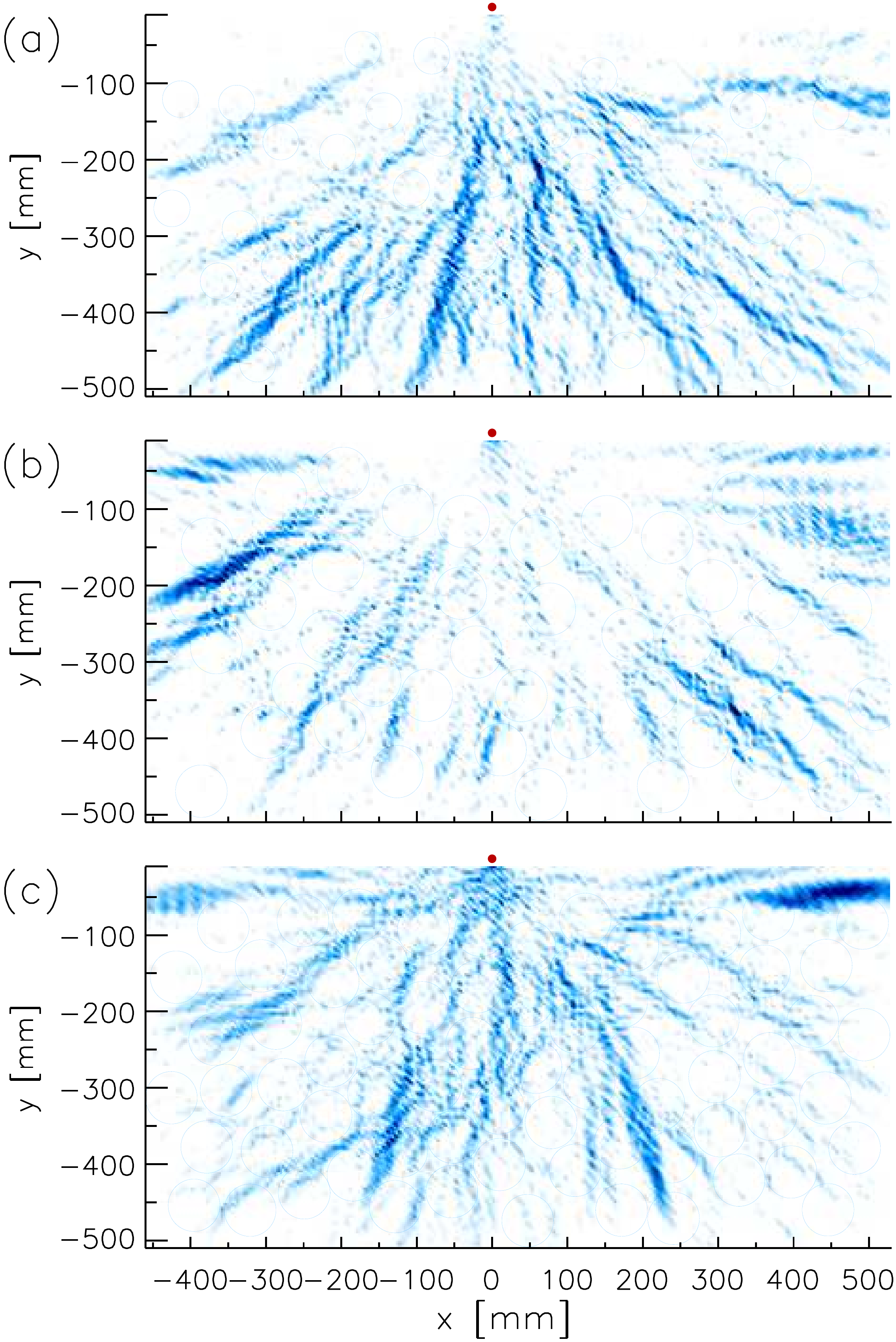}
\caption{\label{fig:ExpFlow} (color online)
Experimental branched flow at 25\,GHz for (a) configuration 1 (caps type 1), (b) configuration~2 (caps type 2), and (c) configuration~3 (caps type 2). The contours of the scatterers are indicated by circles, and a color code from white (low intensities) to blue (high intensities) is used. The source antenna position $r_a$=(0,0)\,mm is marked by a filled circle.}
\end{figure}

Waves propagating through weakly scattering random media surprisingly show pronounced branched intensity fluctuations when the random media are correlated on a length scale larger than the wavelength \cite{top01}. These fluctuations are the source of heavy tails in the intensity distribution of the waves leading to much higher probabilities of the formation of freak waves than expected from random wave models, even for purely linear wave propagation \cite{hel08,hoeh10,yin11,mar12,met13}. Branched flows are a very general phenomenon theoretically predicted for a wide range of systems spanning more than 12 orders in magnitude of spatial scales, from electron waves in a two dimensional electron gas in semiconductors on the micrometer scale to sound propagation in the ocean on length scales of thousands of kilometers \cite{wol01}, and it has been experimentally observed for electrons \cite{top01,jur07,aid07,mar12} and microwaves \cite{hoeh10}. Examples of branched flows in microwave experiments are shown in Fig.~\ref{fig:ExpFlow}. A corresponding detailed description is given later in the experimental part of this Letter.

The basic mechanism of branching is the formation of random caustics in the corresponding trajectory or ray flow \cite{kul82,kap02}. Random caustics are expected to cause power-law tails in the wave intensity distribution \cite{kap02}. These, however, will be observable only when the wavelength is many orders of magnitude smaller than the typical spatial scale of the disorder. For more realistic wavelengths the fingerprint of branching in the intensity distribution of the waves remains an open problem because of the multiple mechanisms influencing this distribution. Thus, while qualitatively the correspondence of theory and experiment is well established, a quantitative comparison allowing to confirm the underlying mechanisms is still missing. One of the most fundamental predictions of theory is the typical scale along the propagation direction at which branching occurs and its scaling with the disorder parameters, adding a new fundamental length scale to transport in random media. In this Letter we first establish, using numerical simulations, that the scintillation index as a function of propagation distance is an adequate quantity to observe the scaling of the branching length. We then present experiments on microwave propagation in quasi-two-dimensional resonators with randomly positioned spherical caps acting as weak scatterers, which confirm the predicted scaling.

\begin{figure*}
\includegraphics[width=1\textwidth]{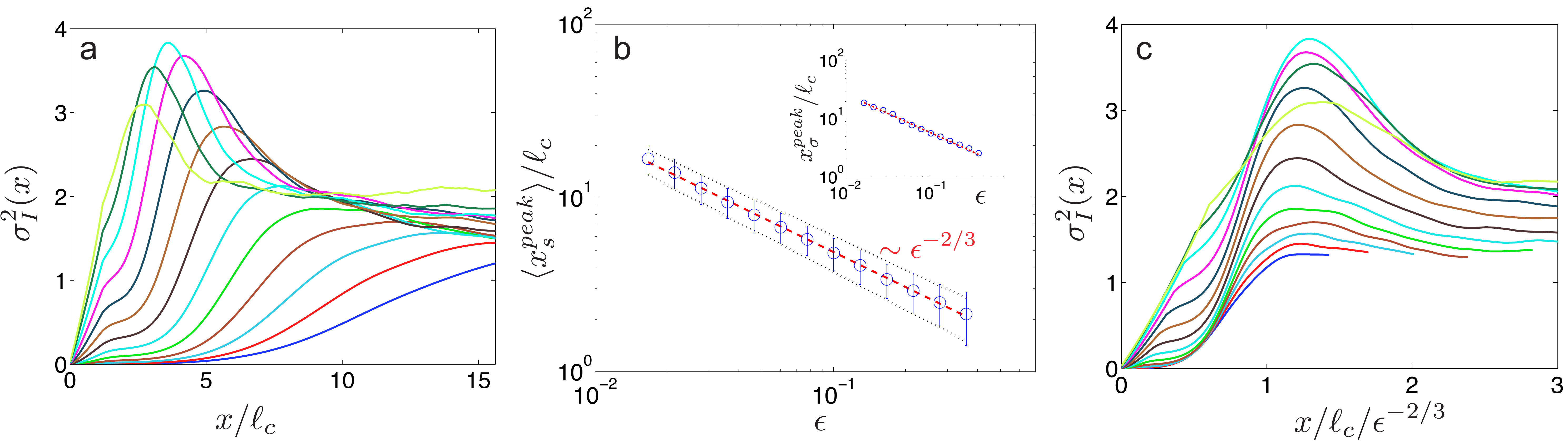}
\caption{\label{fig:numeric_scaling} (color online)
Scaling of the branching length with the strength of the random potential $\epsilon$. (a) Scintillation index $\sigma^2_I(x)$ as a function of the distance from the source for different values of $\epsilon$. (b) Peak positions of the scintillation curves obtained from $s^2_I(x)$ and $\sigma^2_I(x)$ (inset). Both curves show a scaling of $\epsilon^{-2/3}$ (red dashed lines). The black dotted lines indicate the standard deviation of the individual peak positions around their mean value. The scaling is confirmed in panel (c), where the curves from the left panel are shown with a rescaled $x$ axis, on which all peaks occur at approximately the same distance. The peaks of the two curves for the strongest potential, i.e., the two leftmost curves of panel (a), start to decrease in amplitude with growing $\epsilon$, which we attribute to the onset of a significant amount of backscattering.}
\end{figure*}

As a model system, we use the two-dimensional time-independent Schr\"{o}dinger equation which describes the propagation of a wave function $\psi(\mathbf{r})$ in an isotropic disorder potential $V$ which is characterized by its correlation function $c(\mathbf{r})=\langle V(\mathbf{r}^\prime+\mathbf{r})V(\mathbf{r}^\prime)\rangle-\langle V\rangle^2$ where $\mathbf{r}=(x,y)$ is a two-dimensional position vector and the average is taken over realizations of the random potential. Typically a Gaussian correlation is assumed
\begin{equation} \label{eq:Pot_GaussCorrel}
c(|\mathbf{r}|) = \langle V(\mathbf{r}^\prime)V(\mathbf{r}^{\prime}+\mathbf{r})\rangle= \sigma_V^2e^{-|\mathbf{r}|^2/l_c^2}
\end{equation}
where the mean $\langle V(\vec r)\rangle = 0$, $\sigma_V$ denotes the potential strength, and $l_c$ its correlation length. The Gaussian assumption is not necessary but the correlation function should be characterized by two parameters, the standard deviation of the potential $\sigma_V$ and the correlation length $\ell_{c}$. We assume that the potential is weak compared to the kinetic energy of the flow $E$, i.e.~$\epsilon=\sigma_V/E\ll1$,
such that the flow is essentially paraxial and that the wavelength $\lambda$ is smaller than the correlation length, which is necessary for the waves to be able to resolve the structure of the random potential and thus for focusing to occur. The typical distance defining the characteristic length scale of branching is the location at which the wave flow is focused, which leads to the occurrence of extreme waves. This typical distance, $x_{c}$, is most easily analyzed by studying the ray propagation corresponding to the wave flow. Considering a plane wave initial condition and the paraxial approximation valid for a weak random potential, the ray equations can be written in terms of the coordinate transverse to the flow, $y$, as
\[
dy(t)/dt=p_{y}(t),\qquad dp_{y}(t)=-\partial V(t,y)/\partial y
\]
where $t$ now plays the role of the distance along the main flow propagation $x$. Caustics begin to appear in the flow when the rays first start to cross. To calculate the typical distance for a ray to travel to a caustic $x_{c}$, it is thus sufficient to calculate how far the rays have to travel in $x$ to cover approximately one correlation length in $y$ \cite{kap02}, which yields the typical length scale of branched flow,
\begin{equation}\label{eq:x_c}	
  x_{c}/\ell_{c} = \alpha \epsilon^{-2/3},
\end{equation}
where $\alpha$ is a proportionality factor that, e.g., depends on the functional form of the correlation function.
This can also be obtained by more elaborate calculations, and also holds for different initial conditions such as the point source used in our experiments \cite{kul82,kly05,met10} and can also be extended to the case of a perpendicular magnetic field \cite{mar12}. The scaling of $x_{c}$ has been found numerically to deviate from Eq.~\ref{eq:x_c} for $\epsilon \gtrsim 10 \%$, indicating that the paraxial approximation fails and that the potential can no longer be considered to be weak.

The $2/3$ law in the disorder strength described above is a property of the ray propagation, and it is not immediately obvious how this is observable in the wave flow. However, the most striking property of branched flow, namely the high wave intensities caused by diverging ray densities at the caustics, are expected to appear prominently in the intensity statistics of the wave flow. A simple measure to study the branching regime is thus given by the scintillation index (see, e.g., \cite{and99}) as a function of the propagation distance

\begin{equation}
\sigma_I^{2}(x)=\left\langle I^{2}(x)\right\rangle /\left\langle I(x)\right\rangle ^{2}-1, \label{eq:sigma2}
\end{equation}
where $I$ is the wave intensity and the average is taken over realizations of the random potential. We find indeed that a peak in the scintillation index $x_{\sigma}^{\rm peak}$, signals the onset of branching and the occurrence of extreme waves and thus expect its position to obey the same scaling as the typical distance to the first caustics. An alternative measure of the scintillation, which is more appropriate when only a few realizations are available as in the case of our experiments, is to average each individual realization over the transverse coordinate (here $y$) to obtain
\begin{equation}\label{eq:s2}
	s_I^{2}(x)=\left\langle I^{2}(x)\right\rangle_y /\left\langle I(x)\right\rangle_y ^{2}-1.
\end{equation}
One can then determine the peak of these scintillation curves for each realization, and average the resulting peak positions over the realizations of the random potential to obtain $\langle x^{\rm peak}_{s} \rangle$. We expect both scintillation measures to scale according to the $2/3$-law, and confirmed this assumption numerically by propagating a plane wave through 100 realizations of weak random potentials with a Gaussian correlation function for different potential strengths $\epsilon$. As shown in Fig.~\ref{fig:numeric_scaling}, the scintillation curve shows a peak at the onset of branching and a perfect scaling with $\epsilon^{-2/3}$, as expected from the above arguments. Also the scaling of the averaged peaks of $s_I^2(x)$, $\langle x^{\rm peak}_{s} \rangle$, scales as expected. Remarkably, the scaling range extends to $\epsilon\approx 40 \%$ and thus much further than that of $x_c$. We attribute this to the fact, that the peak in the scintillation index originates from the onset of branching, which occurs closer to the source than the mean distance $x_c$ of all rays to have reached the first caustics. The paraxial approximation can thus be expected to hold up to higher $\epsilon$.

%%%%%%%%%%%%%% experimental part %%%%%%%%%%%%%%%%%%%%%%%%%%%%%%%%%%%%

Experimentally we realize the systems by an open microwave cavity~\cite{stoe99}, which makes it possible to measure wave dynamics with a well-controlled tabletop microwave experiment. The metallic top and bottom plates of the cavity are fixed in a distance of $h = 20$\,mm. There are small holes in the top plate on a 5\,mm grid through which a thin wire antenna is inserted extending $3$\,mm into the cavity. A second Teflon-coated antenna is fixed in the bottom plate and extends $5$\,mm into the cavity. Thereby spatially resolved transmission measurements can be automatically performed over an accessible measurement area of $1060 \times 520$\,mm$^2$. The complex transmission spectra $S_{21}$ for each point is measured. This setup was already used for measurements of the Goos-H\"{a}nchen shift \cite{unt11} and of mushroom billiards~\cite{bit10b}. In the pure cylindrical case, i.e., with all side walls parallel to $z$, the vectorial Maxwell equations reduce to a scalar equation for the $z$ component of the electric field. We will neglect here the magnetic field as we will only excite and measure the electric field due to the used dipole antennas. The electric field is thus given by $E_{n,z}(x,y,z) = E_n(x,y)\cos(n\pi z/h)$, where $n$ is the mode number in the $z$ direction. The experiment is described by the two dimensional Helmholtz equation
\begin{figure}[b]
\includegraphics[width=0.95\columnwidth]{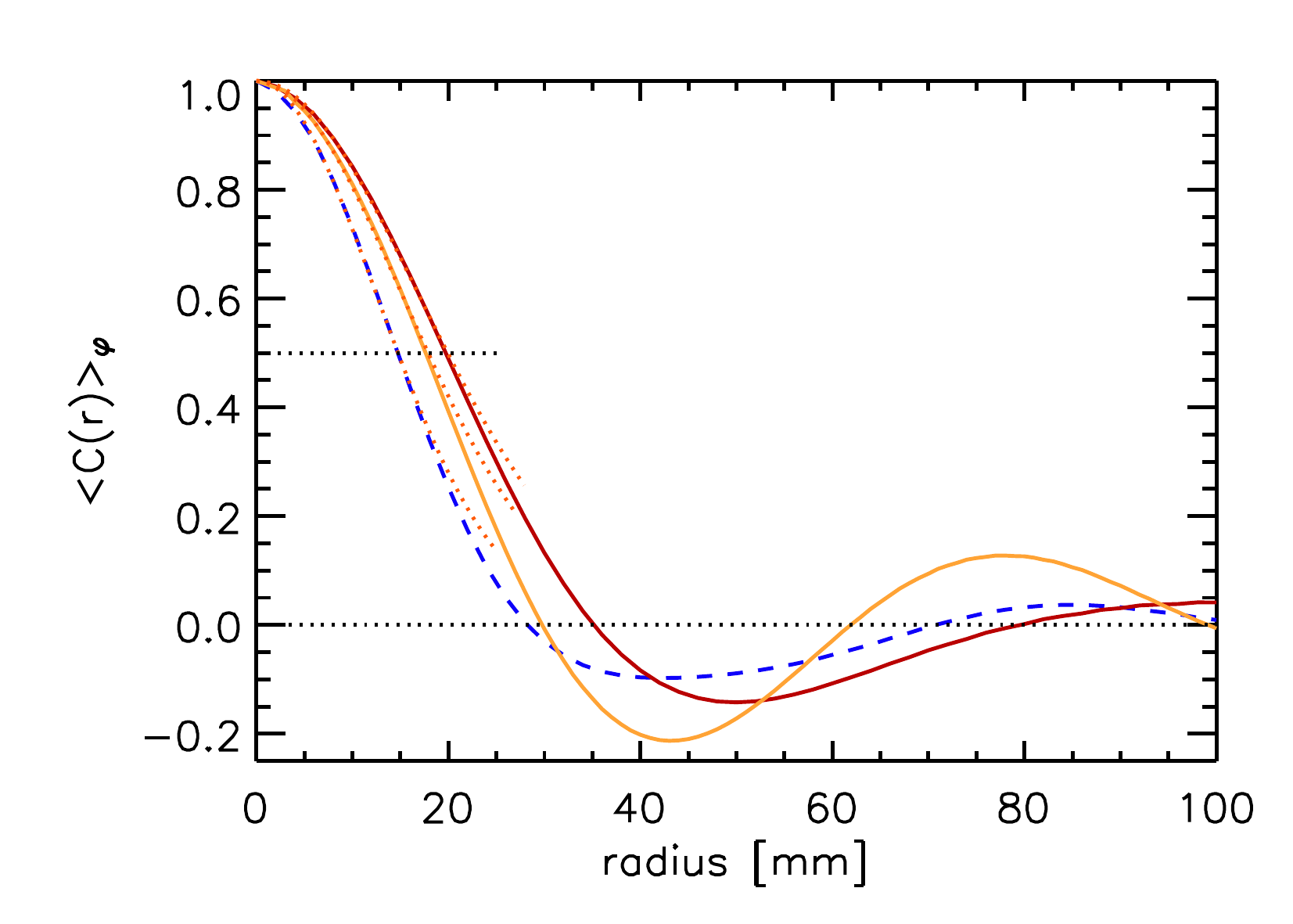}
\hspace*{.33\columnwidth}\raisebox{3.6cm}[0pt][0pt]{\includegraphics[width=.4\columnwidth]{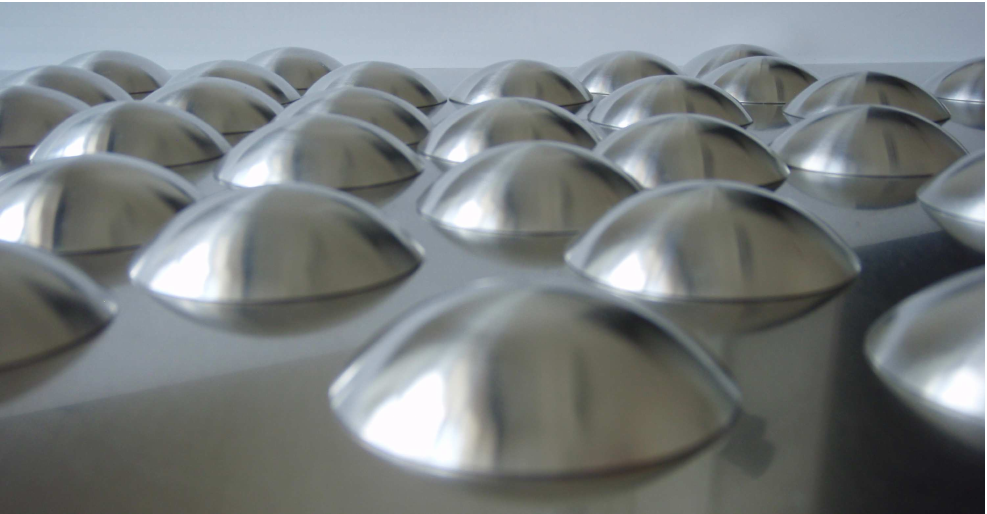}}\\
\caption{\label{fig:Correlfct} (color online)
Autocorrelation function of potential 1 (blue dashed line), 2 (red solid line)  and 3 (orange solid line) with corresponding Gaussian fits (orange dotted lines). For the fits all values above 0.5 (horizontal, black dotted line) were considered. The horizontal dotted line at 0 emphasizes the undershoot of the functions. The inset shows a photo of a configuration with the caps of type I.}
\end{figure}
\begin{eqnarray} \label{eq:Helmholtz_pot}
\left[-\Delta_{xy} + k_{n,z}^2\right]E_{n,z}(x,y)=k_{n,xy}^2E_{n,z}(x,y)
\end{eqnarray}
where $k_{n,z}^2= (n\pi/h)^2$. A comparison with the analogous two-dimensional Schr\"{o}dinger equation
\begin{eqnarray} \nonumber
\frac{\hbar^2}{2m}\left[-\Delta_{xy}+V(x,y)\right]\psi(x,y)=E\psi(x,y)
\end{eqnarray}
suggests to interpret the $k_{n,z}^2$ term as a potential \cite{lau94b,kim05b,hoeh10}. The wave number $k_{n,xy}^2$ corresponds to the quantum mechanical eigenvalue $E$ and can be calculated from $k^2 = k_{n,xy}^2+k_{n,z}^2$. $k$ is the wave number corresponding to the excitation frequency. By varying the height $h(x,y)$ between the two plates we can introduce a potential landscape $V$ for the propagating waves according to
\begin{equation} \label{eq:Potential_Helmh}
V_n(x,y) =  \left(\frac{n\pi}{h(x,y)}\right)^2
\end{equation}
The height variation is realized by distributing spherical caps over the full field, where each cap brings in a central repulsive potential for $n>0$. For the first configuration the spherical caps have a radius $r = 30$\,mm and a height of $0.4 r$ (type I) and for the second and third configurations larger caps with radius $r = 60$\,mm and a height of $0.18 r$ (type II) are used. A photo of the scatterers of type I is shown in the inset of Fig.~\ref{fig:Correlfct}. Previous experiments on freak waves were performed with conical scatterers leading to a singularity of the potential at the cusp, not meeting the requirement of a weak, smooth potential \cite{hoeh10}. But already their typical branching patterns were found in the high frequency regime. The contours of the scatterers are indicated by circles in the wave plots in Fig.~\ref{fig:ExpFlow}.

We measured at frequencies from 15 to 30\,GHz. Thus up to 5 modes, the TM$_0$ to TM$_4$ modes, are open each with different wave number $k_{n,xy}$. For a proper analysis of a single wave with a fixed energy a mode separation by means of a two dimensional spatial Fourier filtering is performed. The found pattern is spatially Fourier transformed and just a ring $k_{1,xy}-0.5(k_{1,xy}-k_{2,xy})\le k \le k_{1,xy}+0.5(k_{0,xy}-k_{1,xy})$ around $k_{1,xy}$ is back transformed. By this procedure we remove all other modes and the signal contains only the TM$_1$ mode. In the further analysis we will concentrate only on this mode. We find an exponential decay of this mode in radial direction, which is removed individually for every frequency, analogously to the procedure in \cite{hoeh10}. The exponential decay comes from the scattering to the TM$_0$ mode. In Fig.~\ref{fig:ExpFlow} examples of measured flows for the three different configurations are shown. We clearly observe here branched flows with different characteristics.

Let us now characterize the different potentials used in the experiment. The standard deviations of the potentials are $\sigma_1 = 19\,951.1$\,m$^{-2}$, $\sigma_2 = 19\,559.5$\,m$^{-2}$ and $\sigma_3 = 23\,488.8$\,m$^{-2}$. To fix the correlation length $l_c$ we calculated the radial correlation function of the potentials. In Fig.~\ref{fig:Correlfct} the correlation function is shown for the three configurations, involving the two scatterer types. The first decay is mainly due to the local potential directly induced by the scatterer, whereas the anticorrelation, e.g., the minimum, is due to the average distance of the scatterer. For the $l_c$ the first decay is the crucial one and we extract $l_c$ by fitting the first decay to a Gaussian [see Eq.~(\ref{eq:Pot_GaussCorrel})]. The extracted correlation lengths for the different configurations are $l_{c,1} = 17.72$\,mm, $l_{c,2} = 23.93$\,mm, and $l_{c,3} = 21.48$\,mm. Note that they are smaller than the radii of the scatterers of $24$ and $34.34$\,mm, respectively. Probably the undershoot caused by the self-avoiding of the scatterers, which cannot be placed overlapping, leads to a compression of the first oscillation. As the scatterer density varies in configurations 2 and 3 the compression effect is stronger in the latter case leading to a steeper decay and a smaller correlation length.

\begin{figure}
\includegraphics[width=0.95\columnwidth]{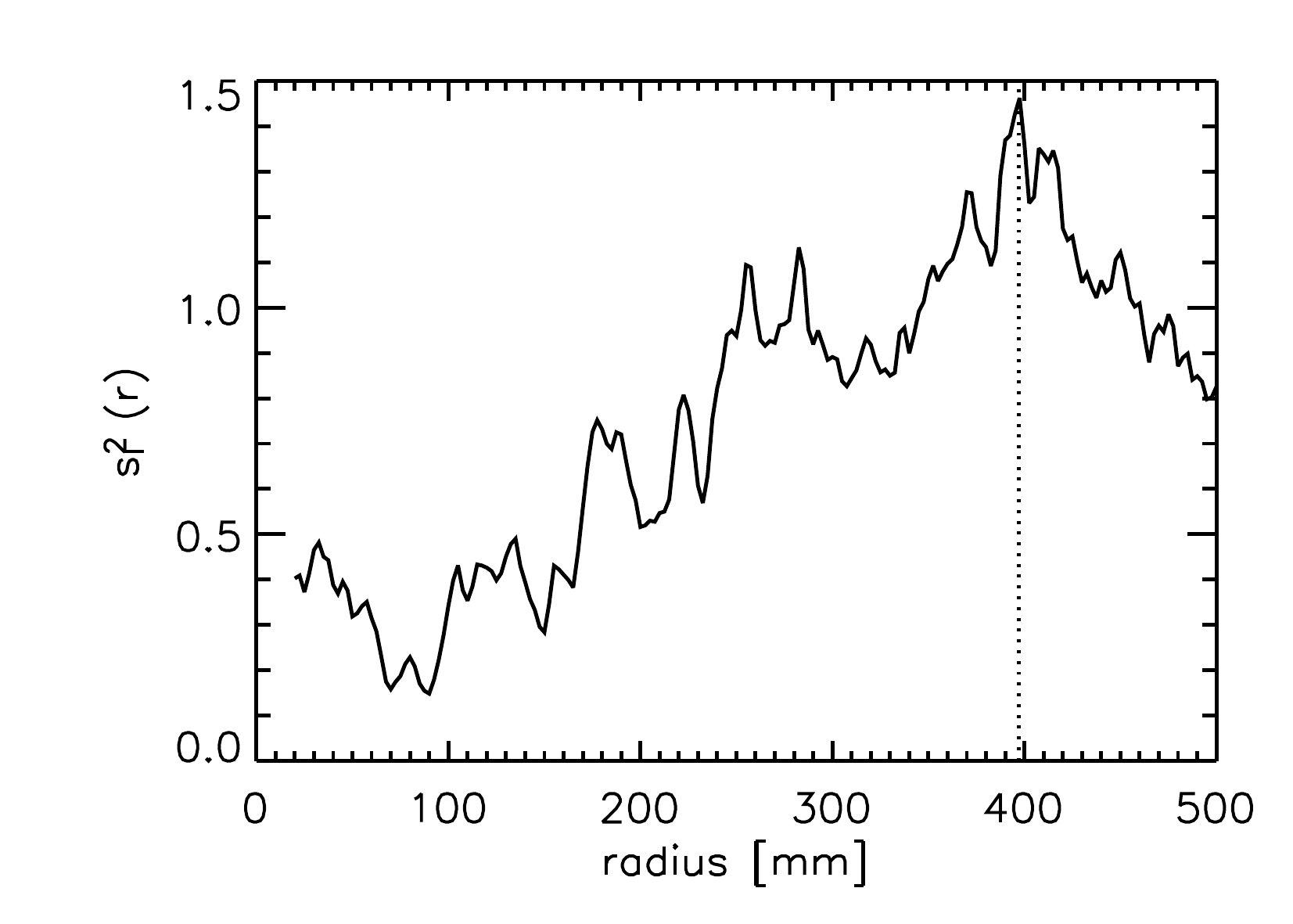}
\caption{\label{fig:var12} (color online)
Scintillation index $s^2(r)$ for configuration 1 at 30\,GHz. The dotted vertical line indicates the extracted maximal position $r_s^{\rm peak}$.}
\end{figure}

From now on all lengths, in particular the radial distance from the antenna $r$, are scaled with the corresponding correlation length. For the extraction of $r_s^{\rm peak}$ we investigate the scintillation index, i.e.~the variance of the flow intensity $A(r,\varphi) = I(r)/\langle I(r) \rangle_\varphi$, on circles with respect to the distance $r$ to the source
\begin{equation}\label{eq:var_circ}
s^2_I(r) = \langle [A(r, \varphi)]^2\rangle_\varphi-1,
\end{equation}
where $\varphi$ is the angular coordinate. Because of the point source the scintillation index now depends on the distance $r$ and not on $x$. The scintillation index at 30\,GHz is shown in Fig.~\ref{fig:var12}. The extracted maximum is marked by the dotted line.

\begin{figure}[b]
\includegraphics[width=0.95\columnwidth]{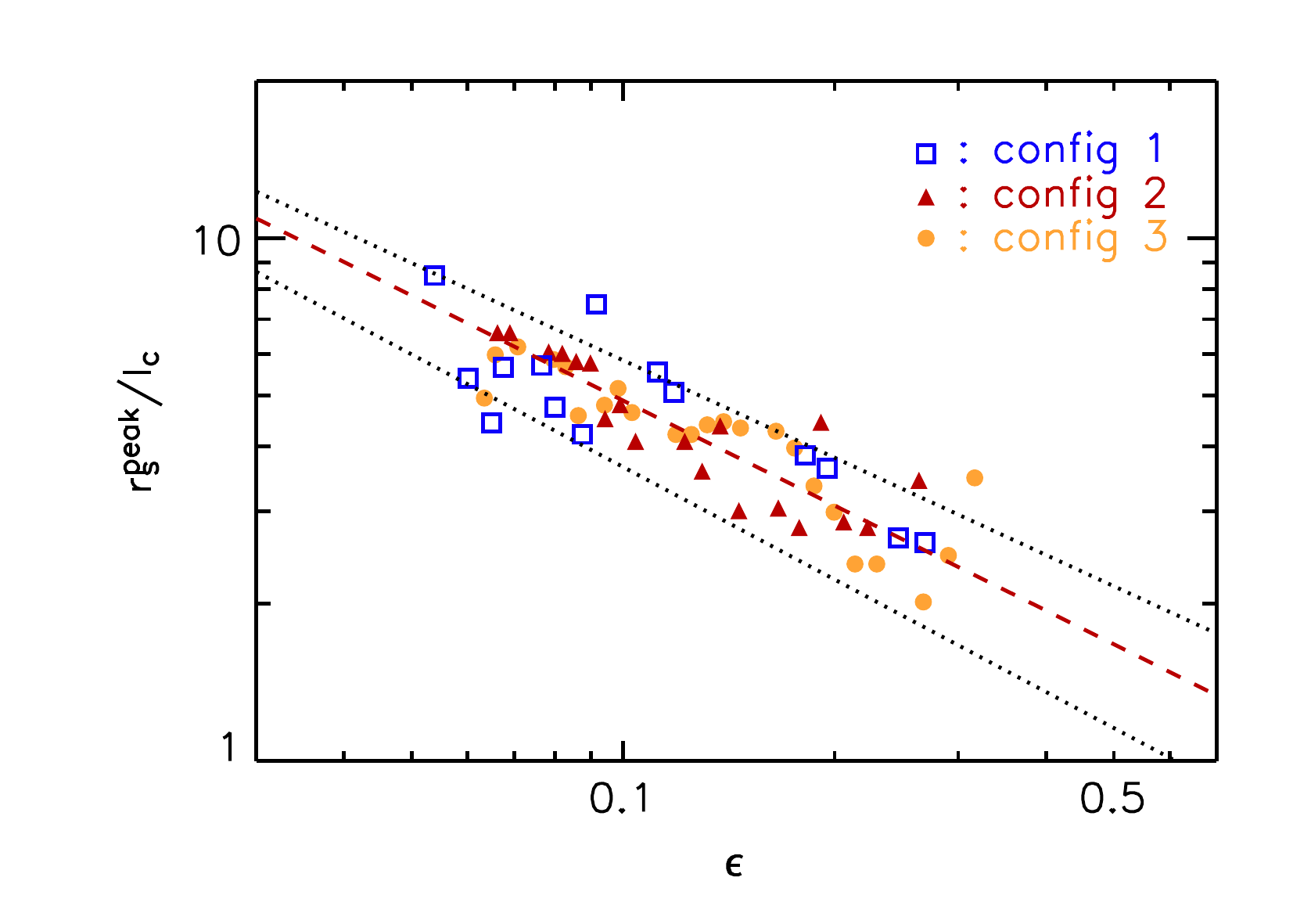}
\caption{\label{fig:scaling} (color online)
Rescaled maxima positions of the scintillation index $r_s^{\rm peak}$ for different frequencies and configurations. The expected scaling behavior of $-2/3$ is indicated by the dashed red line and the standard deviation of the individual peak positions of the simulations (see Fig.~\ref{fig:numeric_scaling} (b)) are marked by the dotted black lines.}
\end{figure}

The maxima of the scintillation index $r_s^{\rm peak}$ for different frequencies and configurations are now extracted. Figure~\ref{fig:scaling} includes the peak positions for all three configurations and different frequencies of the TM$_1$ mode. The peak position is given in units of the correlation length and the standard deviation of the potential is scaled with the corresponding kinetic energy for each frequency, i.e., $\epsilon=\sigma/k_{1,xy}^2$.

The 2/3 scaling, Eq.~\ref{eq:x_c}, includes a proportionality factor which depends nontrivially on the functional form of the correlation function of the random potential and can be calculated if the correlation function is known analytically \cite{kul82,kly05,met10}. Here we adjust the offset to match the values in Fig.~\ref{fig:numeric_scaling}b. The proportionality factors are $\alpha_1 = \exp(1.02)$, $\alpha_2 = \exp(1.12)$, and $\alpha_3 = \exp(1.39)$. We observe that the $\alpha_1$ is similar to $\alpha_2$ as the corresponding configurations have a similar density of scatterers. All three configurations agree with the expected $-2/3$ decay (red dashed line) and their variations are of the same order as the variance of the numerics shown in Fig.~\ref{fig:numeric_scaling} (black dotted lines). Again the clear $-2/3$ decay exceeds the expected limit of validity of $\epsilon \approx 10\%$ by a factor of 3, confirming the above arguments of the very robust paraxial approximation.

In this Letter we have verified the scaling behavior of the branching length numerically and experimentally. We have shown that branching leads to strong intensity fluctuations that are manifested by a peak in the scintillation index, and that its characteristic length can thus be extracted from the scintillation index as a function of propagation distance. We found excellent agreement with the predicted $-2/3$ decay,  even in a much wider range than could be expected from the random caustics statistics.

\begin{acknowledgments}
This work was supported by the Forschergruppe 760 ``Scattering systems with complex dynamics''. S. B. and J. J. M. contributed equally to this work.
\end{acknowledgments}

%%%%%%%%%%%%%%%%%%%%%%%%%%%%%%%%%%%%%%%%%%%%%%%%%%%%%%%%%%%%%%%%%%%%%%%%%%%%%%%%%%%%%%%%%
%merlin.mbs 2010-03-15 4.21a (PWD, AO, DPC)

\end{document}